\documentclass[conference]{IEEEtran}

\usepackage{cite}
\usepackage{amsmath,amssymb,amsfonts}
\usepackage{graphicx}
\usepackage{textcomp}
\usepackage{xcolor}
\usepackage[hyphens]{url}
\usepackage{fancyhdr}

\newcommand{\framework}{Qoncord}
\newcommand{\ignore}[1]{}
\usepackage{enumitem}
\usepackage{multirow}

\usepackage[most]{tcolorbox} %

\usepackage{listings}

\usepackage{balance}
\definecolor{DeepSkyBlue}{HTML}{082567}

\newtcolorbox{highlightbox}[1][]{
  colframe = DeepSkyBlue!100, %
  colback  = DeepSkyBlue!0,  %
  boxsep=1pt, %
  width=\dimexpr\columnwidth\relax, %
  arc=2pt, %
  #1, %
}
\usepackage{graphicx}
\usepackage{textcomp}

\newcommand{\microsubmissionnumber}{348}
\fancypagestyle{firstpage}{
  \fancyhf{}
  
  \fancyhead[C]{\vspace{10pt}\normalsize{MICRO 2024 Submission
      \textbf{\#\microsubmissionnumber} -- Confidential Draft -- Do NOT Distribute!!}\\\vspace{-25pt}} 
  \fancyfoot[C]{\thepage}
}

\renewcommand\IEEEkeywordsname{Keywords}

\title{\framework{}: A Multi-Device Job Scheduling Framework for Variational Quantum Algorithms}

\author{\IEEEauthorblockN{Meng Wang}
\IEEEauthorblockA{\textit{ECE Department} \\
\textit{The University of British Columbia}\\
Vancouver, Canada \\
mengwang@ece.ubc.ca}
\and
\IEEEauthorblockN{Poulami Das}
\IEEEauthorblockA{\textit{ECE Department} \\
\textit{The University of Texas at Austin}\\
Austin, USA \\
poulami.das@utexas.edu}
\and
\IEEEauthorblockN{Prashant J. Nair}
\IEEEauthorblockA{\textit{ECE Department} \\
\textit{The University of British Columbia}\\
Vancouver, Canada \\
prashantnair@ece.ubc.ca}
}

\begin{document}
\maketitle

\pagestyle{plain}

\begin{abstract}

Quantum computers face challenges due to limited resources, particularly in cloud environments. Despite these obstacles, Variational Quantum Algorithms (VQAs) are considered promising applications for present-day Noisy Intermediate-Scale Quantum (NISQ) systems. VQAs require multiple optimization iterations to converge on a globally optimal solution. Moreover, these optimizations, known as restarts, need to be repeated from different points to mitigate the impact of noise. Unfortunately, the job scheduling policies for each VQA task in the cloud are heavily unoptimized. Notably, each VQA execution instance is typically scheduled on a single NISQ device. Given the variety of devices in the cloud, users often prefer higher-fidelity devices to ensure higher-quality solutions. However, this preference leads to increased queueing delays and unbalanced resource utilization.

We propose \textit{\framework{}}, an automated job scheduling framework to address these cloud-centric challenges for VQAs. \framework{} leverages the insight that not all training iterations and restarts are equal, \framework{} strategically divides the training process into exploratory and fine-tuning phases. Early exploratory iterations, more resilient to noise, are executed on less busy machines, while fine-tuning occurs on high-fidelity machines. This adaptive approach mitigates the impact of noise, optimizes resource usage, and reduces queuing delays in cloud environments. \framework{} also significantly reduces execution time and minimizes restart overheads by eliminating low-performance iterations. Thus, \framework{} offers similar solutions 17.4$\times$ faster. It also provides 13.3\% better solutions for the same time budget as the baseline.

\end{abstract}


\begin{figure}[b]
    \parbox{\linewidth}{\vspace{-0.5cm}
    \footnotesize
    \noindent © 2024 IEEE. Personal use of this material is permitted. Permission from IEEE must be obtained for all other uses, in any current or future media, including reprinting/republishing this material for advertising or promotional purposes, creating new collective works, for resale or redistribution to servers or lists, or reuse of any copyrighted component of this work in other works.
    }
\end{figure}

\section{Introduction}
\label{sec:intro}

Quantum computers hold the promise of transformative computational capabilities~\cite{shor1999polynomial,lloyd1996universal,grover1996fast,farhi2014quantum}. While quantum computers with a few hundred qubits are accessible through cloud services from IBM, Amazon, and Microsoft~\cite{IBM_Quantum,AWS_Braket,Azure_Quantum,ibmqroadmap}, their potential is hindered by the inherent hardware noise and scarcity of resources. Near-term \textit{Noisy Intermediate-Scale Quantum (NISQ)} systems~\cite{preskill2018quantum} lack the redundancies necessary for fault tolerance~\cite{micro-taming}. Nevertheless, NISQ devices promise to accelerate many crucial domain-specific applications using \textit{Variational Quantum Algorithms (VQAs)}~\cite{farhi2014quantum,mcclean2016theory,hua2023qasmtrans}. These algorithms map the problems onto parametric quantum circuits and iteratively train circuit parameters with a classical optimizer. However, the performance of VQAs is significantly constrained by noise and prolonged device access times. 

\begin{figure*}
    \centering
    \includegraphics[width=0.9\textwidth]{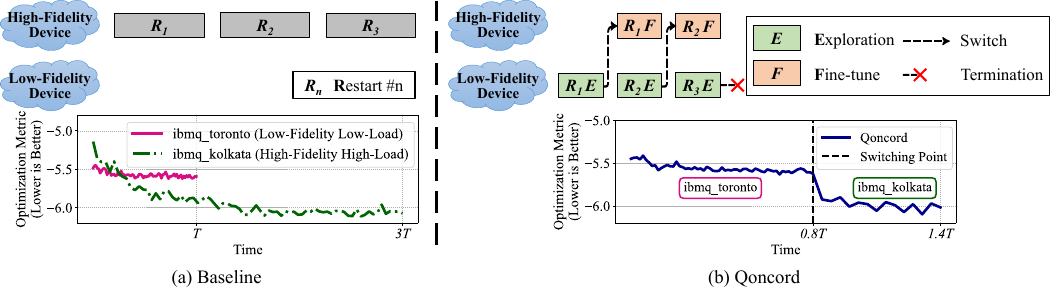}
    \caption{In the baseline approach (Figure (a)), all VQA iterations for each restart are executed on a single device. Despite queueing delays, ibmq\_kolkata outperforms ibmq\_toronto by 159\%, even with 3$\times$ more pending jobs (with longer actual wait times). On the other hand, \framework{} (Figure (b)) starts exploratory iterations on the low-load (LL) device, ibmq\_toronto, before shifting to the high-fidelity ibmq\_kolkata for fine-tuning. This dynamic scheduling leads to 2.14$\times$ faster performance. Additionally, \framework{} efficiently evaluates restart quality on the low-load device, promptly terminating low-quality restarts.}
    \label{fig:intro}
    
\end{figure*}

Program execution on NISQ devices significantly affects the training process of VQAs. For instance, noisy executions can necessitate increased iterations due to a lack of gradients or, in the worst case, not converging to the optimal parameters. The optimization process often gets trapped in local optima, prompting the need for multiple random \textit{restarts}~\cite{redqaoa, graphqaoa}. In these restarts, end-to-end optimization is repeated from a different starting point, and the best result is ultimately selected. Consequently, hardware noise causes the total number of circuit executions required for VQAs to surge compared to an idealized noise-free setting~\cite{tqsim,vqesim,li2024tanqsimtensorcoreacceleratednoisy}.

These problems are exacerbated by (1) the limited availability of quantum computers in the cloud and (2) device-level variations leading to divergent error profiles. Globally, nearly a dozen \textit{quantum cloud services} provide access to fewer than fifty quantum systems~\cite{AWS_Braket,Azure_Quantum,IBM_Quantum,alibaba_cloud,xanadu_cloud,google_quantum_cloud,terra_cloud,dwave_cloud}. These scarce resources are shared among thousands of users, resulting in significant queueing delays. This bottleneck is pronounced for VQAs, which execute circuits sequentially, with hundreds of computational jobs often awaiting execution on a single device~\cite{ravi2021quantum}. Additionally, quantum computers vary in size and error rates. For instance, IBM Quantum systems range from 27 to 127 qubits, with 2-qubit gate error rates varying by 7$\times$ (from 0.3\% to 2.1\%)~\cite{IBM_Quantum}. This leads to unbalanced loads as users and cloud providers naturally select the highest fidelity devices, further complicating resource allocation.

Presently, three approaches exist to mitigate cloud latencies. \textit{Firstly}, IBM's Runtime allows \textit{dedicated} device access to users~\cite{johnson2022qiskit}. Users can specify a device, or the cloud provider defaults to the least busy device based on the load~\cite{qiskitruntime}. \textit{Secondly}, Ravi et al. suggest prioritizing a high-fidelity device with a low load and opting for a low-fidelity device otherwise~\cite{ravi2021quantum}. However, this approach may select low-fidelity devices during periods of high load, potentially compromising solution quality and training performance. \textit{Thirdly}, Stein et al. propose a framework mainly applicable to a class of VQAs for chemistry applications~\cite{stein2022eqc} that require many circuit executions per iteration. The framework distributes the executions across multiple devices (with high and low loads) to accelerate gradient estimation in each iteration. However, all these scheduling policies use the same device(s) \emph{throughout the training}, as shown in Figure~\ref{fig:intro}(a). Thus, their performance is limited by either the noise of the lowest-fidelity device or the wait times of the highest-load device.

Enhancing the performance of VQAs is challenging due to conflicting objectives; devices with \textit{Low-Fidelity} typically exhibit \textit{Low-Load}, whereas \textit{High-Fidelity} devices often experience \textit{High-Load}. Figure~\ref{fig:intro}(a) shows the training steps of a VQA on a noisy simulator using the error model of two 27-qubit IBMQ devices\footnote{It is noteworthy that even with dedicated access to ibmq\_kolkata, users are constrained by access times, especially if numerous users request access.}. We observe that utilizing Low-Fidelity Low-Load devices reduces execution time but compromises accuracy. The ibmq\_kolkata device, with higher fidelity than the ibmq\_toronto device, achieves a 159\% improvement in optimization performance; however, it also encounters prolonged queueing delays, which results in extended execution times. This paper aims to improve throughput (using Low-Fidelity \textbf{Low-Load} devices) without compromising solution quality (using \textbf{High-Fidelity} High-Load devices).

\begin{figure*}
    \centering
    \includegraphics[width=\textwidth]{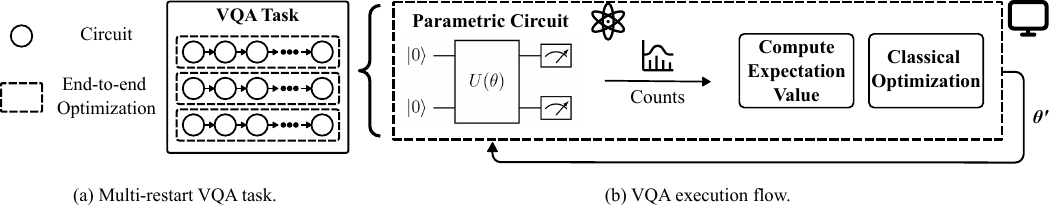}
    \caption{(a) Sample VQA task with three random restarts. (b) VQA executes a parametric circuit whose parameters are tuned over hundreds of iterations by a classical optimization routine. The expectation value from the output distribution is used to update the circuit parameters for the next iteration. }
    \label{fig:background}
\end{figure*}

This paper introduces {\em \framework{}}, an automated job scheduling framework tailored to overcome the above challenges for quantum clouds. \framework{} leverages two insights:

\noindent \textbf{1. Not all Training Steps are Equal:} \framework{} is built on the insight that \textit{not all VQA training steps contribute equally} to the optimization process. It leverages this understanding to recommend that different training steps need not execute on the same quantum device. Thus, \framework{} divides the training process into distinct stages. During the initial \textit{exploration} stage, the optimization routine explores a broad parameter region to identify potential areas containing the optima. The subsequent \textit{fine-tuning} stage involves precisely adjusting the parameter settings to converge to the optimal solution.

Our results, shown in Section~\ref{sec:insight_1}, reveal that the exploration stage is more resilient to noise than the fine-tuning stage. Thus, \framework{} directs the exploration to a Low-Fidelity Low-Load device and dynamically switches to a High-Fidelity High-Load device for performance-critical fine-tuning. As shown in Figure~\ref{fig:intro}(b), this transition occurs if the optimization metric remains unchanged over a specified number (\textit{threshold}) of iterations on the Low-Fidelity Low-Load device. \framework{} achieves fidelity comparable to High-Fidelity-only access in less time by using this scheduling approach.

\noindent \textbf{2. Not all Restarts are Equal:} \framework{} leverages the insight that \textit{not all restarts are equal}, and therefore, not all of them have to go through the end-to-end optimization routine. As illustrated in Figure~\ref{fig:intro}(b), \framework{} quickly evaluates the early exploration steps of each restart on the Low-Fidelity Low-Load device, terminates poor candidates, and fine-tunes the remaining candidates on the High-Fidelity High-Load device. By doing so, the effective throughput of the VQA execution is improved even further. Although Figure~\ref{fig:intro}(b) only illustrates the splitting of the execution into two phases on two devices, \framework{} can be generalized to split the execution into a larger number of phases and then run them on a fleet of NISQ devices with varying fidelities.

\noindent To that end, this paper makes four key contributions:
\begin{enumerate}[leftmargin=0cm,itemindent=.5cm,labelwidth=\itemindent,labelsep=0cm,align=left, listparindent=0.3cm, noitemsep]
    \item We quantitatively demonstrate that high-fidelity quantum devices experience higher loads than other devices.
    \item We show that running all VQA iterations on the same device suffers from poor accuracy or high execution time.  
    \item We propose \framework{}, an automated software framework that splits VQA training into phases and adaptively schedules them across multiple devices. 
    \item \framework{} reduces the overheads of restarts during VQA training by adaptively terminating poor candidates. 
    \item \framework{} provides incentives for device vendors to release non-optimal devices by demonstrating their effective utilization. Thus, it enables cloud service providers to maximize resource utility and reduce queuing delays.
\end{enumerate}

Our evaluations using the error models of IBMQ quantum devices show that \framework{} offers (1)~similar solution quality for VQAs 17.4$\times$ faster and (2)~provides 13.3\% times better solution quality within the same training time. 

\section{Background}

\subsection{Variational Quantum Algorithms (VQAs)}

Variational Quantum Algorithms (VQAs) promise to accelerate diverse applications in healthcare~\cite{ibmqproteinfolding}, combinatorial optimizations~\cite{farhi2014quantum}, quantum chemistry~\cite{mcclean2016theory,peruzzo2014variational}, and machine learning~\cite{biamonte2017quantum} using NISQ devices. These algorithms utilize a parameterized quantum circuit, as illustrated in Figure~\ref{fig:background}. The circuit's gate parameters undergo training via a classical optimizer, optimizing an objective function corresponding to the specific problem over numerous iterations.

The expectation value is computed from the output distribution in each iteration. This metric guides the adjustment of circuit parameters for the subsequent iteration. The training process continues until the optimal parameters are identified. At this convergence point, the circuit's output corresponds to the optimal solution for the given problem.

\subsection{Impact of Quantum Hardware Errors on VQAs}

Noisy quantum hardware hampers VQA performance by extending training time and hindering convergence due to lack of gradients~\cite{mcclean2018barren}. Instead of arriving at any global optima, the optimization process can get trapped in a local optima. To overcome this, VQAs employ a \textit{ multi-restart} strategy that executes independent optimization iterations with different initial parameters, as shown in Figure~\ref{fig:background}. This increases the chances of discovering the global optimum with the cost of additional circuit executions. The number of restarts, typically a few hundred, depends on the circuit complexity~\cite{shaydulin2021qaoakit}.

\subsection{Quantum Computing in the Cloud}

Operating quantum computers requires specialized infrastructures and high-precision control electronics. These requirements incur substantial deployment and operational costs. To address these challenges, \textit{Quantum Cloud Services (QCS)} provides a vital interface that shields users from the complexities of quantum hardware and facilitates seamless remote access. Major device manufacturers, including IBM, Rigetti, and IonQ, offer direct access to their quantum processing units (QPUs) through cloud services. Cloud service providers like Amazon Web Services (AWS) and Microsoft Azure enable access to third-party QPUs. 

\subsection{Demand vs. Supply Gap In Quantum Clouds}

Enterprises such as Boeing, JSR, Exxon, Mitsubishi Chemical, Daimler, CERN, and others are exploring quantum applications through partnerships with QCS providers~\cite{boeing_ibmq,jsr_ibmq,exxon_ibmq,mitsubishi_chemical_ibmq,mitsubishi_ibmq,daimler_ibmq,cern_ibmq,azure_partners,aws_braket_partners,accenture_quantum,xanadu_multiverse,quantinuum_bmw,quantinuum_news}. IBM Quantum alone collaborates with 210+ organizations, spanning Fortune 500 companies, universities, and startups~\cite{IBM_Quantum}. 

Despite this growth, the demand for quantum resources significantly outpaces availability. Globally, only a dozen QCS providers offer access to less than 50 machines~\cite{AWS_Braket,Azure_Quantum,IBM_Quantum,alibaba_cloud,xanadu_cloud,google_quantum_cloud,terra_cloud,dwave_cloud}. This considerable gap between demand and supply cannot be bridged solely by building more quantum computers due to their high operational and maintenance expenses. Consequently, QCS providers must intelligently schedule their limited quantum resources among various users.

\subsection{Job Scheduling Policies}

Access to quantum cloud follows two models: \textit{Shared} and \textit{Runtime}. Both models employ fair-share scheduling, ensuring equitable distribution of computation time among users.

\noindent \textbf{1. Shared Access Model:}
In this model, users submit individual jobs, which are then queued for execution on the quantum device. The fair-share scheduling algorithm determines a job's position in the queue. It considers factors such as the number of requests, the requested computation time, and the user's past usage to allocate resources.

\noindent \textbf{2. Runtime Access Model:}
The runtime model~\cite{johnson2022qiskit} offers a more dynamic interaction with quantum resources. Users establish a continuous session upon gaining access to a quantum device. Users can submit multiple jobs once a session is established without rejoining the queue. Jobs within an active session receive higher priority for execution. This benefits VQAs by reducing latency between iterations. Although the runtime model facilitates a more efficient submission process during an active session, it does not alleviate the underlying queuing challenge. This is because the demand for quantum devices still exceeds supply. Thus, jobs from other users still need to wait while the runtime model prioritizes a user.

These access models overcome the limitations of a single machine chosen to execute all program iterations. Their scheduling algorithm does not consider inter-machine characteristics. Some QCS providers are pursuing intelligent models to counter this. For instance, IBM's recent video for their runtime model envisions a future where the jobs are distributed over multiple machines. However, to our knowledge, no such scheduling algorithm is currently available~\cite{qiskitruntimevideo}.

\section{Motivation}

\subsection{Balancing Loads and Fidelity in Quantum Clouds}

QCS encounters extreme load imbalances. These are primarily driven by variations in device-level error characteristics within diverse quantum computers. Typically, high-fidelity machines experience heavier loads due to the natural inclination to run applications on these devices, creating an imbalance with respect to low-fidelity systems. For instance, Table~\ref{tab:ionq_queue_time} shows the average wait times for different machines from Rigetti and IonQ. The number of algorithmic qubits (\#AQ) is a system-level metric used to benchmark the performance of IonQ machines, and a higher value is desirable~\cite{ionq_aq,chen2023benchmarking,quantumvolume}. This value is unavailable for providers like Rigetti. Thus, we also compare the average 2-qubit gate fidelities.

\begin{table}[h]
    \centering
    \setlength{\tabcolsep}{6pt}
\caption{Fidelity and Wait Times Comparison of Devices}
  
\label{tab:ionq_queue_time}
\resizebox{.9\columnwidth}{!}{
    \begin{tabular}{|c|c|c|c|c|} \hline 
  \multirow{2}{*}{\textbf{Provider}} & \multirow{2}{*}{\textbf{Device}} &  \textbf{Gate} & \multirow{2}{*}{\textbf{\#AQ}} & \textbf{Wait} \\ 
   
   & & \textbf{Fidelity (\%)} & & \textbf{Time}\\ \hline 
   Rigetti~\cite{rigetti_qcs}& Aspen-M-3 & 94.6 & - & 4 hours \\\hline
   \multirow{3}{*}{IonQ~\cite{ionq_cloud}}       & Harmony& 97.1 & 25 & 1.9 days \\ \cline{2-5} 
         & Aria& 98.9 & 25 & 10.7 days\\ \cline{2-5} 
         & Forte& 99.4 & 29 & 7 days\\ \hline
    \end{tabular}}
\end{table}

Notably, the wait times for noisier Rigetti machines are 10.9$\times$ to 61.3$\times$ lower than those for high-fidelity IonQ machines. There is variability even within the same provider (IonQ). Aria and Forte, with higher fidelity, exhibit 3.7$\times$ to 5.6$\times$ longer wait times than Harmony. This unique trade-off presents QCS providers with a choice: either prioritize faster time to solution on low-fidelity devices at the expense of solution quality or endure prolonged wait times for high-fidelity machines to achieve better solution quality.

\subsection{Variance in Capabilities of Quantum Systems}
Quantum systems exhibit a trade-off between fidelity and program execution time due to three fundamental reasons:
\subsubsection{Different qubit device technologies (systems from multiple vendors)}
\label{sec:diff_qubits}
Multiple qubit technologies, including superconducting qubits, trapped ions, and neutral atoms, are being explored to build large-scale quantum computers. Each device technology is associated with its unique execution and error characteristics. They also mature at different rates due to the unique challenges associated with scaling each qubit technology. As shown in Table~\ref{tab:ionq_queue_time}, IonQ's trapped ion devices offer higher fidelity but are more than 1000$\times$ slower (measured as time per gate) than Rigetti's superconducting qubits. Thus, program execution is orders of magnitude slower on IonQ systems compared to Rigetti, even if cloud access latencies are considered to be negligible (assuming user reserves the machine). The problem worsens for VQAs, considered the most promising near-term quantum applications, as they run each program for thousands of iterations, where each iteration is executed for thousands of trials, and the whole iterative process is repeated a few thousand times by starting with new initial program parameters (restarts).  

\begin{table}[htp]
\centering
\caption{Amazon Braket Pricing}
\resizebox{\columnwidth}{!}{
\begin{tabular}{|c|c|c|c|c|} \hline 
\textbf{Provider} & \textbf{Device} & \textbf{Execution Time/Gate} &  \textbf{Price/Task}  & \textbf{Price/Shot}  \\ \hline 
IonQ & Harmony & 200 microseconds & \$0.3 & \$0.01  \\ \hline 
IonQ & Aria  &  600 microseconds & \$0.3& \$0.03  \\ \hline 
IonQ & Forte & 970 microseconds & \$0.3 & \$0.03 \\  \hline 
Rigetti & Aspen-M & 169 nanoseconds  & \$0.3 & \$0.00035 \\ \hline
\end{tabular}
}
\label{tab:aws_cost}
\end{table}

The trade-off in performance metrics is visible in the competitive cost dynamics for accessing these resources. Table~\ref{tab:aws_cost} shows the cost of accessing various quantum systems on AWS Braket. Users incur an access cost to initiate a task and a shot cost based on the number of shots and per-shot price. Typically, applications run for thousands of shots. On AWS Braket, the higher-fidelity but slower Aria system costs 3$\times$ more per shot than the lower-fidelity and faster Harmony~\cite{ionq_cloud}. More noticeably, Rigetti offers more competitive pricing, with per-shot rates 28.6$\times$ to 85.7$\times$ lower than IonQ, making them attractive alternatives to users, if they can be used efficiently.

\subsubsection{Different systems from the same provider (same qubit technology)}
\label{sec:same_vendor}
Systems from the same vendor using identical device technology can exhibit significant variability in error rates and gate durations. For instance, IonQ-Forte, the latest device from IonQ, offers higher fidelity but slower gate times compared to its predecessors, Aria and Harmony~\cite{ionq_cloud}. Similarly, ibmq\_kolkata demonstrates lower noise levels than ibmq\_toronto, which features an older architectural design~\cite{IBM_Quantum}. Even within the same fabrication technology, process variation leads to differences in error characteristics. For example, two Eagle processors, ibm\_kyiv and ibm\_cleveland, have 1.5\% and 6.6\% errors per layered gate, respectively~\cite{IBM_Quantum}.

\subsubsection{Single system with different software capabilities}
\label{sec:single_device}
\begin{figure}
    \centering
    \includegraphics{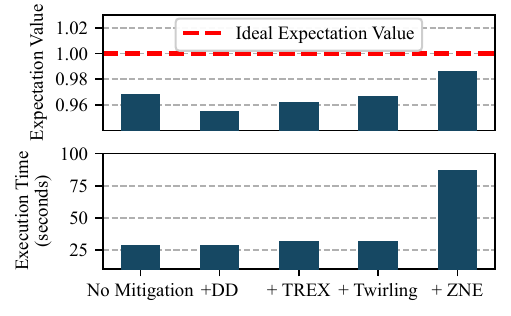}
    \caption{Trade-offs between fidelity (expectation values) and execution latency for a 50-qubit two-local ansatz on the 127-qubit ibm\_kyoto device, using five error mitigation modes: no mitigation, dynamic decoupling (DD), twirled readout error extinction (TREX), gate twirling, and zero noise extrapolation (ZNE). Error mitigation improves fidelity at the cost of increased latency.}
    \label{fig:error_mitiq} 
\end{figure}

Execution results can vary significantly even on the same system depending on the software capabilities employed. For example, the quantum volume (a benchmarking measure for quantum systems) of ibmq\_montreal increased from 64~\cite{jurcevic2021demonstration} to 128~\cite{ibm_quantum_volume} within 6 months by only using software error mitigation. Error mitigation techniques, such as Dynamic Decoupling (DD)~\cite{viola1998dynamical}, Twirled readout error extinction (TREX)~\cite{maciejewski2020mitigation, bravyi2021mitigating}, Gate Twirling~\cite{emerson2007symmetrized}, and Zero Noise Extrapolation (ZNE)~\cite{temme2017error}, can substantially improve the fidelity of programs but often come at the cost of increased execution time and computational overheads. For example, as shown in Figure~\ref{fig:error_mitiq}, applying ZNE to a 50-qubit two-local ansatz on the ibm\_kyoto device reduces the error by 57-70\% but also increases the execution time by 3$\times$. This shows that program execution must navigate the fidelity versus execution time trade-off efficiently to achieve both high accuracy and low execution time. %

The fundamental trade-off between fidelity and total execution latency worsens at the cloud level as multiple users attempt to navigate them at the application level, resulting in prolonged wait times. 
\framework{} optimizes for both fidelity and execution time through enhanced job scheduling. \framework{} improves the performance of VQAs when systems of multiple qubit technologies (Section~\ref{sec:diff_qubits}), as well as systems of the same qubit technology, are used (Section~\ref{sec:same_vendor}). 

\ignore{
\framework{} demonstrates remarkable versatility, effectively addressing the challenges posed by diverse qubit technologies, multiple generations of systems from the same vendor, and varying software capabilities on individual devices. Moreover, it presents a compelling case for device vendors to maintain and offer lower-fidelity quantum devices alongside their cutting-edge counterparts. By leveraging \framework{}'s ability to utilize a spectrum of quantum resources efficiently, vendors can maximize the utility and lifespan of their entire device portfolio, potentially reducing costs and broadening access to quantum computing resources. 
}

\ignore{ 
The pricing and performance of cloud quantum computers from major providers reveal key tradeoffs impacting user adoption. Amazon Braket's pricing for gate-model quantum services from IonQ, OQC, and Rigetti highlights the complex balance between cost and fidelity. Each provider's pricing structure reflects their strategic positioning - balancing expense with qubit quality to target different market segments. As the quantum computing landscape matures, analyzing providers' pricing decisions and resulting customer behaviors will illuminate the interplay between price, performance, and demand.
}

\section{\framework{}}
\label{section: design}

This paper introduces {\em \framework{}}, a fully automated job scheduling software for NISQ applications on quantum clouds.

\subsection{Design Philosophy}

To mitigate queuing latency, loads from a high-load device can be distributed to a low-load device. However, fully migrating a Variational Quantum Algorithm (VQA) task from a high-load device to a low-load device can introduce fidelity concerns. Typically, low-load devices exhibit lower queuing latencies due to their lower fidelity.

In response, we designed \framework{} based on the observation that VQA tasks can be divided into two distinct phases, each with varying resilience to noise. The noise-resilient phase is efficiently queued onto low-fidelity devices, and only the noise-sensitive phase is executed on the high-fidelity device. This effectively reduces the overall queuing time for the VQA task and maintains the same solution quality as if the task were executed entirely on the high-fidelity device.

\subsection{Insight 1: Not All VQA Iterations are Equal}
\label{sec:insight_1}

\begin{figure}[b]
    \centering
    
    \includegraphics[width=0.9\columnwidth]{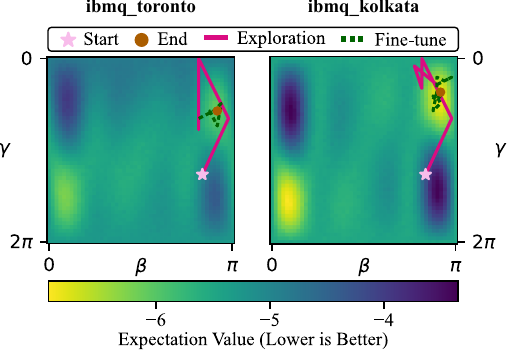}
    \caption{Landscape of a 7-qubit VQA and the optimizer path traced during training on two 27-qubit IBMQ systems. }
    \label{fig:noisy_landscape}
\end{figure}

The VQA training process navigates the optimization landscape to converge on optimal parameter values. For example, in Figure~\ref{fig:noisy_landscape}, the training path is depicted on the landscape of a 7-qubit VQA benchmark using a noisy simulator. This path aims to achieve optimal parameter values for $\beta$ and $\gamma$. The error model of two 27-qubit IBMQ machines, namely ibmq\_toronto and ibmq\_kolkata, is used for this study. We start the training process with the same initial parameter values for both machines to highlight our insight.

The parameters $\beta$ and $\gamma$ are tunable and are modified with each iteration using gradient descent-based classical optimizers. The gradient descent operation is employed on an optimization landscape with several symmetric optimal regions; Figure~\ref{fig:noisy_landscape} shows two such symmetric optimal regions, namely one at the upper right and one at the lower left.

\noindent\textbf{Bimodel Phases:} The early training phase, known as the \textit{exploration} phase, distinguishes between regions with and without optima. In this phase, the trajectory moves from the sub-optimal region in the bottom right to the optimal upper right region. Once the region with the optima is identified, a more precise parameter \textit{fine-tuning} (shown by the dotted line), called the fine-tuning phase, locates the optima.

\noindent\textbf{Observation:} Relying on gradient values within the optimization landscape can enable us to identify the end of the exploration phase and the beginning of the fine-tuning phase. For instance, in Figure~\ref{fig:noisy_landscape}, we observe that:

\begin{enumerate}[leftmargin=*, labelsep=0.1cm,align=left,labelwidth=\itemindent, listparindent=0.3cm, noitemsep]
    \item There exist gradients on the optimization landscape for both machines. The gradients tend to saturate while the VQA task executes on the lower-fidelity ibmq\_toronto device. This can point to the end of the exploration phase.

    \item In the exploration phase, the optimization tends to proceed in the same direction on lower-fidelity ibmq\_toronto and higher-fidelity ibmq\_kolkata. Thus, this phase can be executed accurately, even on a low-fidelity device.
    
    \item  As the exploration phase concludes, the gradients tend to be sharper on the higher-fidelity ibmq\_kolkata machine. This sharper gradient landscape is sufficient to distinguish between the regions with and without the optima -- essentially identifying and triggering the fine-tuning phase. 
    
    \item On the contrary, on the ibmq\_toronto device, the gradients are insufficient for fine-tuning. As a result, the optimization process does not converge on this device. In contrast, the fine-tuning phase is successful on ibmq\_kolkata.
\end{enumerate}

\noindent \textbf{Approach:} Starting on an inexpensive, low-latency, and lower-fidelity quantum device allows us to explore the search space quickly. We can then switch to expensive, high-latency, high-fidelity devices for fine-tuning. This can maximize resource utilization and enable faster time-to-solution.

\subsection{Insight 2: Not All Restarts Are Equal}~\label{notallrestarts}

\begin{figure}[h]
    \centering
    \includegraphics[width=0.9\columnwidth]{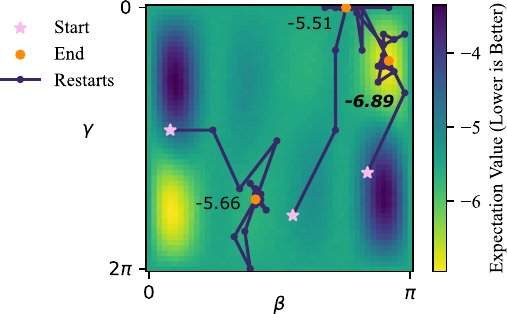}
    \caption{Optimization paths for three restarts starting from unique initial points on the landscape of a 7-qubit VQA. Only one of the restarts converges to the global optimum (corresponding to the expectation value -6.89).}
    \label{fig:restarts}
\end{figure}

When training a VQA task, multiple rounds of optimization are required. This is because the optimization landscape can have many local optimal points in addition to the global optimum. If the training process only goes through a single round, it risks converging at one of the local optimal points instead of finding the best global solution. 

\noindent\textbf{Mitigating Useless Restarts:} To overcome this issue, multiple training rounds or restarts are conducted, each with a distinct set of initial parameters. This approach allows each restart to explore a different region of the optimization landscape. Figure~\ref{fig:restarts} illustrates this method for three separate restarts from different initial points on the landscape of a 7-qubit VQA. Even in this relatively low-dimensional space (with only two parameters), only \emph{one} restart converges to the global optima. Typically, large circuits require as many as a hundred restarts, and our experiments observed that only 40\% achieved convergence to a global optimum. Furthermore, the number of restarts increases as the number of parameters increases, posing significant computational overhead~\cite{farhi2014quantum,peruzzo2014variational,mcclean2016theory,yuan2019theory}.

\begin{figure}[h]
    \centering
    \includegraphics[width=0.9\linewidth]{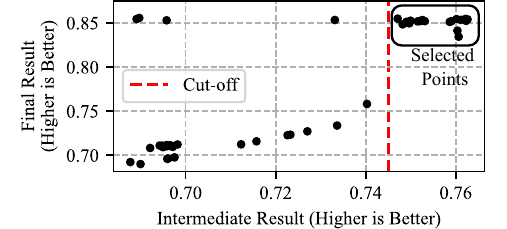}
    \caption{"Scatter plot comparing the final result (after executing all iterations) with the intermediate result (after completing 40\% of the iterations). Typically, an optimization that converges to a global optimal value also demonstrates good performance in the early stages of optimization.}
    \label{fig:clustered_results}
\end{figure}

\begin{figure*}
    \centering
    \includegraphics[width=0.9\linewidth]{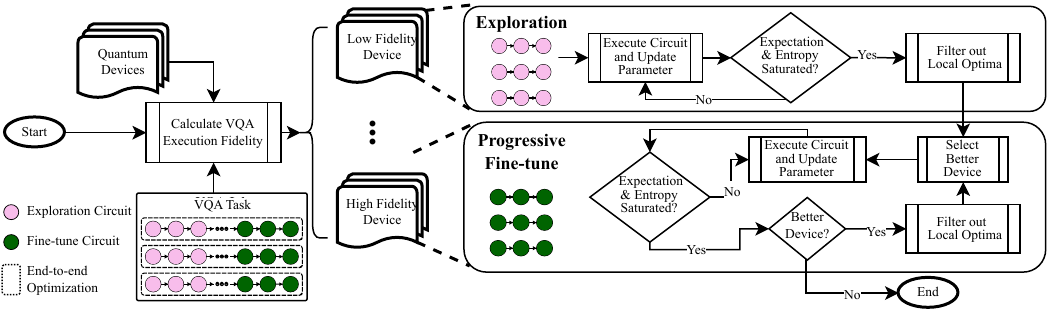}
    \caption{Overview of \framework{}: It computes the expected execution fidelity of a given VQA task on all available devices. Then, it starts the \textit{exploration} stage on the low-fidelity device for all restarts. Once optimization terminates, \framework{} migrates to a high-fidelity device for \textit{fine-tuning}. }
    \label{fig:design_overview} %
\end{figure*}
\noindent\textbf{Intermediate Value Clusters:} One can identify high-quality restarts by observing their intermediate values. Figure~\ref{fig:clustered_results} shows that, across various VQA circuits, the intermediate expectation values of high-quality restarts tend to be clustered. These clusters are formed by executing only 40\% of the iterations per restart and lie in the exploration phase.

\noindent \textbf{Approach:} \framework{} reduces the restart overhead by quickly evaluating the quality of each restart using the low-fidelity, low-load device. This enables \framework{} to \emph{only} fine-tune the high-quality restarts on high-fidelity devices. 

\ignore{
As the number of parameters in the VQA increases, the search space dimension grows exponentially. Unfortunately, current VQAs require parameterized quantum circuits with many parameters to achieve the quantum advantage over classical algorithms \cite{farhi2014quantum,peruzzo2014variational,mcclean2016theory,yuan2019theory}. Each additional restart introduces a complete end-to-end optimization that further exacerbates the already substantial computational burden of VQA executions.}

\subsection{Implementation Overview}

Until now, we've emphasized dividing the VQA optimization into exploration and fine-tuning phases. However, a more crucial realization is that each quantum device has inherent noise, limiting its VQA optimization capabilities. Devices with lower noise and higher fidelity enhance the optimization process, leading to better solutions. Therefore, \framework{} surpasses static iteration distribution across devices.

The implementation overview of \framework{} is shown in Figure~\ref{fig:design_overview}. Two key components facilitate dynamic scheduling in \framework{}: 1) An execution fidelity estimator that predicts the impact of device noise and ranks expected performances across devices. 2) An adaptive convergence checker setting termination conditions per device based on metrics such as expectation value and Shannon entropy.

As shown in Figure~\ref{fig:design_overview}, \framework{} starts with the fidelity estimator to select appropriate devices for the given VQA task, then initiates exploration on the lowest-fidelity device for all restarts. The adaptive convergence checker monitors each restart, terminating low-quality restarts with intermediate values not close to a clustered set. Only high-quality restarts progress to the next higher-fidelity device for continued exploration and fine-tuning. This iterative process advances the device hierarchy toward maximum fidelity.

\subsection{Execution Fidelity Estimator}
\label{sec:fidelity_estimator}

The initial step in \framework{} is to evaluate the fidelities of available quantum devices for various stages of the optimization process. By default, \framework{} employs a metric, denoted as $P_{\text{Correct}}$, measuring the probability of obtaining a correct outcome on a given device. This metric, derived from prior work~\cite{stein2022eqc}, is determined using Equation~\eqref{eq:p_correct}:

\begin{equation}
P_{\text{Correct}} = e^{-\frac{CD \frac{\mu_{tG_1}+\mu_{tG_2}}{2} }{T_1T_2}}(1 - \gamma)^{G_1}(1 - \beta)^{G_2}(1 - \omega)^M
\label{eq:p_correct}
\end{equation}

\noindent where $CD$ represents circuit depth, $\mu_{tG_1}$ and $\mu_{tG_2}$ denote average latencies for single and two-qubit gates, respectively. The fidelity of single-qubit gates, two-qubit gates, and measurements is represented by $\gamma$, $\beta$, and $\omega$, respectively. $G_{1/2}$ and $M$ indicate the number of single and two-qubit gates and the measurements count, respectively.

\begin{figure}[h]
    \centering
    \includegraphics[width=0.9\linewidth]{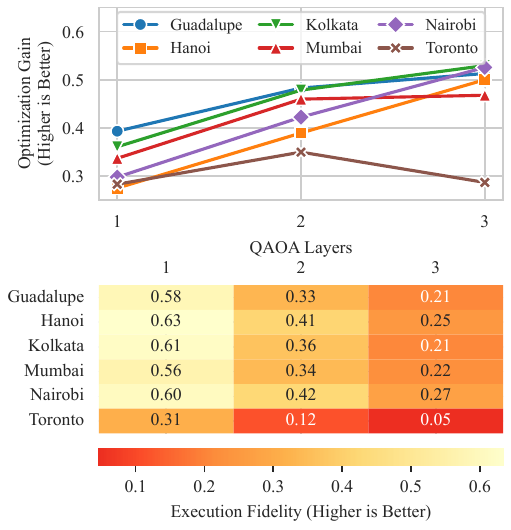}
    \caption{Optimization performance of a 7-qubit QAOA using noise models from six IBMQ devices with up to three layers. While more layers enable better solutions (in theory), they are more error-prone on real systems due to additional gates. The heatmap shows the estimated circuit fidelity calculated using Equation~\ref{eq:p_correct}. Fidelity below 0.1 gives poor results.}
    \label{fig:minimum_p_correct}
\end{figure}

\noindent\textbf{Minimum Estimated Fidelity:} Given a VQA task and a set of available devices, \framework{} initially filters out devices that are too noisy to run the task effectively. This process involves determining a minimum acceptable circuit fidelity, as calculated in Equation~\ref{eq:p_correct}. In Figure \ref{fig:minimum_p_correct}, a sample 7-qubit QAOA circuit is simulated using noise models from six IBM quantum devices. The optimization gain reported is the increase in the approximation ratio of QAOA achieved through classical optimization. While, in theory, more QAOA layers enable finding better solutions~\cite{farhi2014quantum}, i.e. with a higher optimization gain, additional layers also introduce more gates, leading to increased execution errors. The heatmap illustrates the estimated fidelity per device and layer count. It becomes evident that below an estimated fidelity of 0.1, adding more QAOA layers ceases to improve performance due to excessive noise, indicating a plateau in optimization progress. To address this, \framework{} establishes a minimum fidelity threshold of \textbf{0.1}. Device-task combinations with estimated fidelities below this threshold are excluded for that VQA, ensuring backend hardware can meaningfully optimize for the given task.

\subsection{Adaptive Convergence Checker}
\label{sec:convergence_check}

$P_{\text{Correct}}$ offers a high-level understanding of the overall execution error rate for a specific VQA task on a given quantum device. However, this alone proves insufficient for effective VQA task scheduling across multiple devices. While $P_\text{Correct}$ provides valuable insights into the estimated execution fidelity of the quantum circuit running on a given device, it fails to fully capture the progress of the optimization process. This is primarily due to the fact that these metrics do not consider that parameters in the ansatz circuit of VQAs can cause variance in the final execution fidelity.

\begin{figure}[h]
    \centering
    \includegraphics[width=0.9\linewidth]{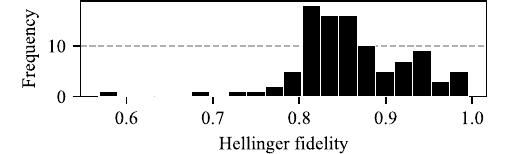}
    \caption{Distribution of the Hellinger fidelity of a 7-qubit, 1-layer QAOA circuit running with 100 random sets of parameters. Different parameter values have different levels of noise tolerance, thus, resulting in different fidelity values.}
    \label{fig:fidelity_diverge}
\end{figure}

\noindent\textbf{Why $P_{\text{Correct}}$ may not indicate progress:} Consider the 7-qubit QAOA circuit from Figure~\ref{fig:minimum_p_correct} as an example. When executing on the ibmq\_kolkata device with 1 QAOA layer, $P_{\text{Correct}}=0.61$ is calculated. However, as shown in Figure~\ref{fig:fidelity_diverge}, running the same circuit with 100 random parameter sets results in varying Hellinger fidelities~\cite{hellinger1909neue} (from 0.56 to 0.99). The calculated $P_{\text{Correct}}$ does not align with the mean Hellinger fidelity of 0.83 which was recorded. This indicates $P_{\text{Correct}}$ cannot provide enough context on the current optimization progress as it's insensitive to parameter changes.

\noindent\textbf{Shannon's Entropy versus Quality of Results:} The Shannon entropy~\cite{shannon1948mathematical} of the output distribution serves as another valuable metric to assess optimization progress. Entropy provides insights into the optimization trajectory beyond the expectation value. A high entropy value signals greater output randomness and more average-case results. On the other hand, a low entropy value indicates less uncertainty and is typically associated with either optimized or worst-case results.

\begin{figure}[h]
    \centering
    \includegraphics[width=0.9\linewidth]{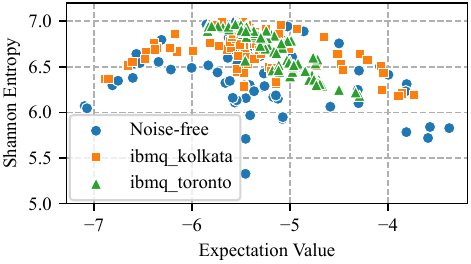}
    \caption{Relationship between Shannon entropy and expectation values for a 7-qubit QAOA benchmark. Entropy generally anti-correlates with expectation value, but the relationship is complex. The noisy ibmq\_toronto progresses from low to higher entropy states but does not converge. In contrast, the higher-fidelity ibmq\_kolkata better resolves the characteristic entropy arc.}
    \label{fig:entropy}
\end{figure}

\noindent\textbf{Assess Progress using Joint Entropy and Expectation:} As the VQA optimization converges, one would expect entropy to evolve from initially low (potentially a worst-case starting point) and transition through higher entropy average-case distributions before decreasing again as superior solutions are approached. Figure~\ref{fig:entropy} visually presents this entropy arc, partially captured on noisy devices like ibmq\_toronto. In contrast, higher-fidelity ibmq\_kolkata demonstrates progression towards an apparent better solution. As shown in Figure~\ref{fig:entropy}, the same entropy value can correlate to different expectation values, and vice versa. This non-linear relationship can be exploited to avoid premature optimization termination as a single measure might indicate that no further optimization is necessary while the other still shows potential for improvement. Therefore, \framework{} utilizes both entropy and expectation values to ensure that the optimization only terminates when both metrics have stabilized at their optimal levels. When transitioning to higher-fidelity systems, \framework{} checks if entropy decreases, indicating less noisy execution, to determine whether to stay on the current device or move to the next tier.

\subsection{Optimization Strategy in \framework{}}

In \framework{}, optimizations unfold across a series of devices instead of completing end-to-end on a single machine. Before reaching the final device for fine-tuning, \framework{} employs a relaxed convergence checker with less strict criteria than the final convergence decision. For instance, if the original checker terminates optimization after \emph{ten} iterations without improvement, the relaxed checker may instead trigger it at \emph{five} iterations. This accounts for further exploration that may enhance the solution in subsequent stages. Only on the final device is the stringent original convergence checker applied to decide when to terminate the VQA ultimately. This two-tiered strategy prevents premature convergence while reducing unnecessary iterations on noisier devices.

\subsection{Efficient Restart Selection}
VQAs typically require multiple random restarts to avoid local optima~\cite{shaydulin2021qaoakit}. To this end, \framework{} selectively chooses restarts to proceed to higher-quality hardware for fine-tuning. Leveraging the insight presented in Section~\ref{notallrestarts}, \framework{} assesses the quality of restarts by examining their intermediate expectation values. Only restarts found within the top-performing cluster, as determined by this analysis, are promoted to downstream devices. Conversely, restarts deemed of low quality are terminated, enhancing overall efficiency.

\subsection{Maintaining Up-To-Date Calibration}
Quantum device calibration data in Equation~\ref{eq:p_correct} can become outdated. Thorough calibrations on devices are expensive and infrequent \cite{knill2008randomized, ibm_calibration, ravi2021quantum, stein2022eqc}. Cloud service providers could address this transparently by periodically storing a sample of optimization outcomes and comparing new outcomes to these benchmarks. In this way, drifts in device noise profiles can be detected without added executions.

\section{Methodology}

\subsection{Cloud Scheduling Policy}
\label{sec:schduling}
We compare \framework{} against the following schedulers:
\begin{enumerate}[leftmargin=0cm,itemindent=.5cm,labelwidth=\itemindent,labelsep=0cm,align=left, listparindent=0.3cm, itemsep=5pt]
    
    \item \textbf{Least Busy:}~\cite{ravi2021quantum,IBM_Quantum} This policy always selects the least occupied device for job execution, theoretically offering the highest throughput but resulting in a degraded accuracy. 

    \item \textbf{Load Weighted:} Devices are chosen based on their load, with less loaded machines being more likely to be selected.
    
    \item \textbf{Fidelity Weighted:} Devices are selected based on their fidelity, which is the general user access patterns.

    \item \textbf{Best Fidelity:}~\cite{ravi2021quantum} This policy always selects one of the highest fidelity devices, aiming for the best possible results but potentially incurring longer latencies due to the limited number of high-performance devices.

    \item \textbf{EQC:}~\cite{stein2022eqc} This policy involves the ensemble execution of variational quantum algorithms across all devices. It increases the total number of circuits to be executed. In our evaluation, the EQC scheduling is modeled by converting runtime jobs into individual tasks and using the least-busy method for scheduling. Each runtime job under this model requires twice the number of tasks compared to other methods, representing the minimum overhead for a 1-layer QAOA.
\end{enumerate}

\begin{highlightbox}
Ideally, the wait time of the \textit{Least Busy} machine and fidelity of the \textit{Best Fidelity} machine is desired.
\end{highlightbox}

\subsection{Figure of Merit}
The evaluation focuses on system throughput and accuracy.

\noindent\textbf{1. Throughput:} Throughput is the number of tasks completed per unit of time. To evaluate throughput in our study, we calculate it using the formula:
\begin{equation}
\label{eq:throughput_calculation}
\text{Throughput} = \frac{\text{Number of Circuits}}{\text{Completion Time}}
\end{equation}
This metric allows us to compare the efficiency of different scheduling methods under identical conditions. It is important to note that for the EQC policy, the number of circuits is higher relative to other policies. A detailed discussion on EQC scheduling is provided in Section~\ref{sec:eqc}.

\noindent\textbf{2. Accuracy:} To assess accuracy, we evaluate the performance of VQAs using \textit{approximation ratio}. It compares the optimized expectation value ($E_{\text{optimized}}$) derived from the VQA to the ground truth expectation value ($E_{\text{ground\_truth}}$) obtained through brute force searching and is the theoretical minimum expectation value. Mathematically, this ratio is expressed as:
\begin{equation}
\label{eq:approximation_ratio}
\text{Approximation Ratio} = \frac{E_{\text{optimized}}}{E_{\text{ground\_truth}}}
\end{equation}

\begin{figure}[h!]
    \centering
    \includegraphics[width=0.7\linewidth]{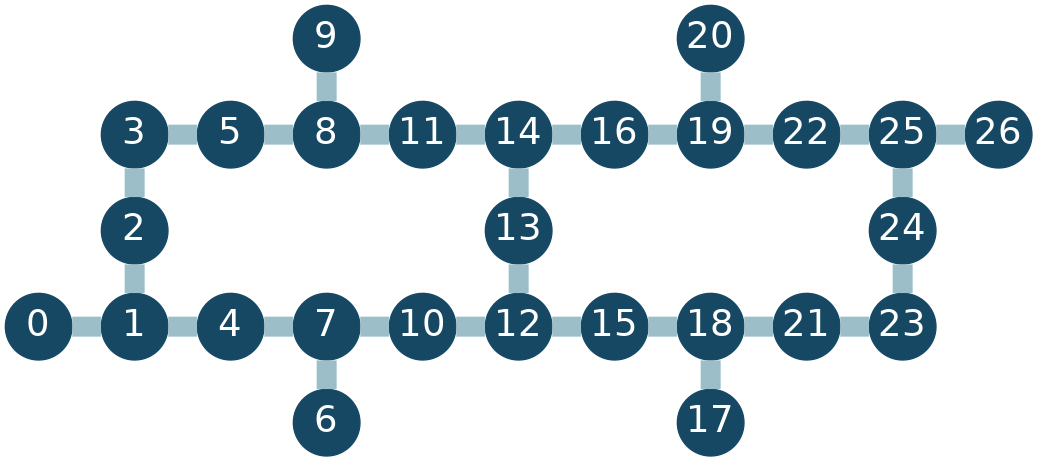}
    \caption{Coupling map of the 27-qubit IBM devices.}
    \label{fig:coupling_map}
\end{figure}

\subsection{VQA Algorithms}
We assess \framework{} using two prominent VQAs: Quantum Approximated Optimization Algorithm (QAOA)~\cite{farhi2014quantum} and Variational Quantum Eigensolver (VQE)~\cite{peruzzo2014variational}. For QAOA, we focus on the max-cut problem with two Erdős Rényi random graphs~\cite{erdds1959random}: one with seven and the other with nine nodes. Both graphs are generated with an edge probability of 0.5. We examine QAOA circuits with 1, 2, and 3 layers. For VQE, we choose the hydrogen molecule ($H_2$) and use a 4-qubit unitary coupled cluster single and double excitation (UCCSD)~\cite{bartlett2007coupled} ansatz to find the ground state energy of $H_2$. All the circuits are transpiled using qiskit transpiler with O3 optimization.

\subsection{Noisy Simulation Setup}
By default, we use device error profiles from ibmq\_kolkata and ibmq\_toronto. As shown in Figure~\ref{fig:coupling_map}, both devices have the same coupling map. ibmq\_kolkata (high-fidelity device) has an average 2-qubit gate error rate of 1.091\% and readout error rate of 1.22\%. ibmq\_toronto (low-fidelity device), on the other hand, has a higher average 2-qubit gate error rate of 2.083\% and readout error rate of 4.48\%. The 36-qubit IonQ-Forte device noise profile is also used for sensitivity studies. It has an all-to-all connection with an average 2-qubit gate error of 0.74\% and an average readout error of 0.5\%.

\subsection{Execution Platform}
We use Qiskit~\cite{Qiskit} 0.45.0 for the circuit simulation and optimization trajectory analysis. We use noise models generated from ibmq\_toronto, ibmq\_kolkata, and IonQ-
Forte for noisy simulations. Circuit simulations are run on a cluster with a 40-core Intel Xeon Gold 6230 processor nodes (2.1GHz with 192GB DDR4-2666 ECC memory). We use Qiskit's implementation of the Simultaneous Perturbation Stochastic Approximation (SPSA)~\cite{robbins1951stochastic} for the classical optimizer.

\subsection{Scheduling Simulation Setup}
We compare scheduling policies discussed in Section~\ref{sec:schduling}. Our experiment aims to replicate real-world conditions and involves a pseudo workload of 1000 quantum jobs. These jobs include independent tasks, executed once, and runtime jobs, which account for 10 to 90\% of all jobs, that continuously submit circuit executions during an active session. Variable delays separate the consecutive executions within a runtime session to mimic the behavior of runtime job scheduling in real-world workloads ~\cite{johnson2022qiskit}. This allowed for the insertion of other queued jobs. Execution times randomly vary 3× between minimum and maximum, reflecting empirical hardware behavior~\cite{ravi2021quantum}. Tests cover hardware with an execution fidelity between 0.3–0.9.

\section{Results and Analysis}

\subsection{Performance Analysis of Scheduling Policies}

\begin{figure}[h!]
    \centering
    \includegraphics[width=0.95\linewidth]{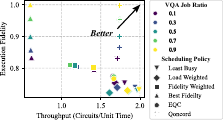}
    \caption{Fidelity-throuput analysis. We evaluate different scheduling policies against Qoncord using a simulation of 1000 quantum jobs on 10 hypothetical devices with fidelities ranging from 0.3 to 0.9. Qoncord is the closest to an ideal policy as all its points lie closer to the top right-hand corner.}
    \label{fig:schdeuling_simulation}
\end{figure}

Figure~\ref{fig:schdeuling_simulation} shows the fidelity relative to the highest-fidelity device and execution throughput. \framework{} consistently delivers high fidelity while having a high throughput that is very close to the \texttt{Least Busy} policy even when the device fidelity gap alters due to temporal variations in error-rates. framework{} outperforms other policies as the percentage of VQA jobs or total cloud load increases. This capability is essential for VQAs, which demand both accuracy and efficiency. It also enables cloud providers to reduce the time to access machines (and time to solution) without compromising accuracy. Policies such as \texttt{Least Busy} and \texttt{Load Balanced} achieve higher fidelities but also significantly compromise fidelity. The \texttt{Fidelity Weighted} policy does not excel in either metric. EQC scheduling, despite operating on a least busy principle, faces challenges due to the 2$\times$ circuit execution (discussed more in Section~\ref{sec:eqc}) overhead of VQA tasks. This overhead resulted in only moderate improvements in throughput. Note that a direct assessment of the average fidelity of EQC is not applicable because, unlike other scheduling strategies, including \framework{}, EQC employs an \emph{asynchronous} gradient descent strategy. This aspect of EQC is discussed more in detail in Section~\ref{sec:eqc}.

\subsection{End-to-End Multi-Restart VQA Optimization}
\label{sec:res:multi_restart}

\begin{figure}[h]
    \centering
    \includegraphics[width=0.9\linewidth]{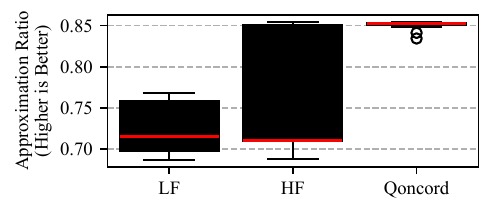}
    \caption{Distribution of approximation ratios across 50 random QAOA optimization restarts. \framework{} starts optimization on the LF device and fine-tunes on the HF device. Overall, \framework{} matches the maximum approximation ratio achieved of HF-only optimization and its result, on average, is at least 20\% higher than any of the single-device optimization results. } 
    \label{fig:multi_restart_approx}
\end{figure}
We evaluate the impact of \framework{} on restarts required for VQAs. Figure \ref{fig:multi_restart_approx} shows the approximation ratios achieved by \framework{}, the HF device alone, and the LF device alone across 50 random restarts for a 3-layer QAOA benchmark. \framework{} filters out 31 bad-performing optimization runs when transitioning from LF to HF device and only the remaining 19 subsequently progressed on the HF device. As a result, \framework{} produces a mean approximation ratio over 20\% higher than the other cases.

\begin{figure}[htp]
    \centering
    \includegraphics[width=0.9\linewidth]{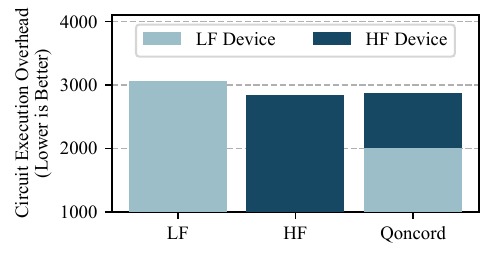}
    \caption{Circuit execution overhead for 50 random QAOA restarts across three execution modes: LF-only, HF-only, and \framework{}. All three modes incur similar overall circuit execution overheads. However, \framework{} distributes its executions evenly across available devices, this greatly optimizes the overall resource usage and enhances execution efficiency.}
    \label{fig:multi-restarts-nfev}
\end{figure}

While providing a result with higher quality, \framework{} also reduces the load on each device. Figure \ref{fig:multi-restarts-nfev} shows the circuit execution overheads of using LF-device only, HF-device only, and \framework{}. In this example, using only the LF device requires 3,055 circuit executions, whereas using the HF device alone requires 2,841 circuit executions. \framework{}, on the other hand, requires a total of 2,865 circuit executions, out of which 2,009 are executed on the LF device and the remaining 856 on the HF device. This means the LF-device executes 70\% of the total executions while also ensuring greater accuracy due to early termination of poor optimization runs. 

\subsection{Multi-Restart VQA Optimization: More Quantum Devices}

\begin{figure}[h!]
    \centering
    \includegraphics[width=0.9\linewidth]{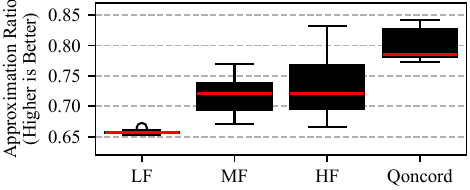}
    \caption{Distribution of approximation ratios across 50 random QAOA optimization restarts. Three device noise profiles are included, ibmq\_toronto is used as low fidelity (LF), ibmq\_kolkata medium fidelity (MF), and IonQ-Forte high fidelity (HF). \framework{} starts optimization on the LF device and progressively moves to higher fidelity devices to enhance the execution result.} 
    \label{fig:multi_restart_opte_3devices}
\end{figure}

\begin{figure}[h!]
    \centering
    \includegraphics[width=0.9\linewidth]{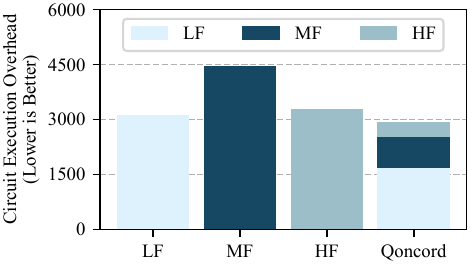}
    \caption{Circuit execution overhead for 50 random QAOA restarts across four execution modes: LF-only, MF-only, HF-only, and \framework{}. LF and HF modes, along with \framework{}, show similar, lower overheads as optimizers quickly converge in very noisy or nearly noise-free environments. In contrast, the MF-only mode exhibits higher overheads, this suggests that with moderate noise levels, significant improvements are possible but harder to achieve.} 
    \label{fig:multi_restart_nfev_3devices}
\end{figure}

\framework{} is designed to integrate additional devices into its execution seamlessly and can iteratively enhance the outcomes of multi-restart optimizations performed with them. We have incorporated three device profiles for demonstration and used them to optimize a 9-qubit, 3-layer QAOA circuit. Specifically, ibmq\_toronto is used as low fidelity (LF), ibmq\_kolkata medium fidelity (MF), and IonQ-Forte high fidelity (HF). Figure~\ref{fig:multi_restart_opte_3devices} shows the distribution of 50 random restarts of the VQA task performed individually on each device and with \framework{}. Notably, \framework{} provides the highest approximation ratio achieved. Furthermore, the average performance of \framework{} across all restarts also significantly exceeds that of any single-device setup, with a more than 8\% improvement in average approximation ratio over all single-device executions.

Figure~\ref{fig:multi_restart_nfev_3devices} shows the corresponding circuit execution overheads. Overall, LF, HF, and \framework{} modes exhibit similar, lower execution overheads. This indicates that \framework{} effectively leverages the available devices by balancing the computational load, thus reducing the demand for higher-fidelity devices. In contrast, the MF-only mode shows significantly higher execution overheads, suggesting that moderate noise levels present a more challenging optimization landscape. 

In essence, \framework{} not only matches the performance of high-fidelity devices but does so with enhanced efficiency by distributing executions in a way that leverages the strengths of each device. This strategic management allows it to maintain high accuracy without placing undue strain on any single device, particularly those with higher fidelity.

\subsection{Multi-Restart VQA Optimization: Larger Quantum Circuits}

\begin{figure}[h!]
    \centering
     \includegraphics{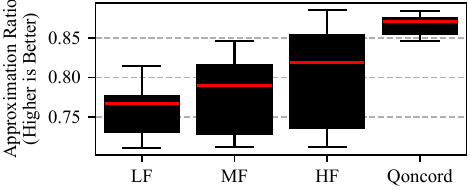}
    \caption{Distribution of approximation ratios across 50 random QAOA optimization restarts for a 14-qubit 1-layer QAOA task.}
    \label{fig:14q_exps}
\end{figure}

\begin{figure}[h!]
    \centering
    \includegraphics{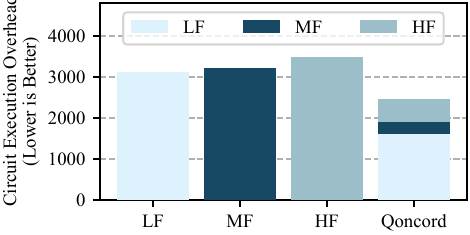}
    \caption{Circuit execution overhead for the 14-qubit QAOA optimization task.}
    \label{fig:14q_nfevs}
\end{figure}

We tested \framework{} on a 14-qubit VQA job, the largest simulatable using density matrix-based noisy quantum circuit simulators on modern GPUs. As no existing quantum device noise profiles met our minimum fidelity criteria (Section~\ref{sec:fidelity_estimator}), we created three hypothetical noise models with depolarization error rates for 2-qubit gates and readout: 0.1\% (high-fidelity), 0.5\% (mid-fidelity), and 1\% (low-fidelity). Figure~\ref{fig:14q_exps} shows approximation ratios for these models and \framework{}. Even at this increased complexity, \framework{} outperformed single-device results. Figure~\ref{fig:14q_nfevs} displays corresponding circuit execution overheads. \framework{} effectively utilized low and mid-fidelity devices while showing better results.

\subsection{Single Restart QAOA Optimization}

\begin{figure}[h]
    \centering
    \includegraphics[width=0.9\linewidth]{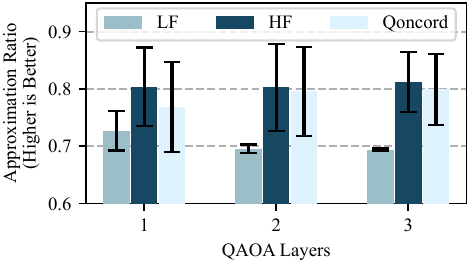}
    \caption{Approximation ratio for a single QAOA optimization restart. \framework{} achieves similar performance as the HF-only case.}
   
    \label{fig:single_restart_qaoa}
\end{figure}

We also analyzed the performance of \framework{} on a single VQA optimization restart \emph{without} early termination. We used a 3-layer QAOA circuit and compared it against running the full optimization simulated using noise profiles from the ibmq\_kolkata (HF) and ibmq\_toronto (LF) devices. In theory, running solely on the HF device should produce the best approximation ratio. Our goal with \framework{} is to achieve a comparable result as the HF device while reducing the load placed on that device. Figure \ref{fig:single_restart_qaoa} shows that \framework{} achieves an approximation ratio very close to the HF-only case and over 14\% higher than the LF-only optimization. 

\begin{figure}[h]
    \centering
    \includegraphics[width=0.7\columnwidth]{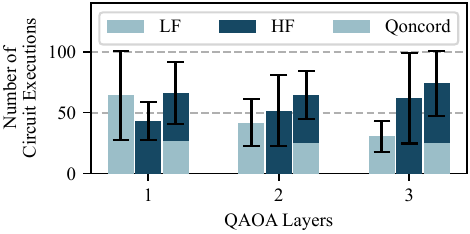}
    \caption{Average number of circuit executions for a single QAOA restart comparison. While the total executions are similar, \framework{} reduces the executions on the individual HF and LF devices.}
    \label{fig:single_restart_execs}
\end{figure}
Furthermore, by dynamically switching between the HF and LF executions, \framework{} reduces the load on each individual device. As seen in Figure \ref{fig:single_restart_execs}, \framework{} required similar total executions as the HF and LF simulations but with fewer executions on either simulated device alone. By balancing computations across devices, \framework{} can achieve results comparable to an HF-only optimization with a lower peak load. This shows the capability to maintain accuracy while mitigating bottlenecks when scaling to larger optimizations.

\subsection{VQA Optimization of VQE}
\begin{figure}[h]
    \centering
    \includegraphics[width=0.8\linewidth]{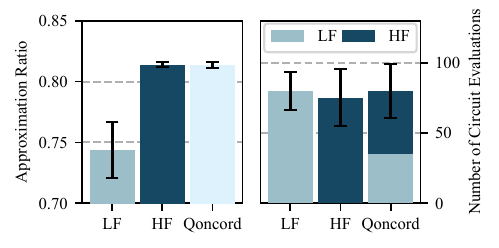}
    \caption{Accuracy and execution overheads for a 4-qubit VQE running with hydrogen molecule and UCCSD ansatz. \framework{} matches HF accuracy with no additional executions beyond those needed for HF or LF optimizations.}
    \label{fig:vqe_results}
\end{figure}

We evaluate \framework{} on a VQE application using a 4-qubit UCCSD ansatz to find the ground state energy of a hydrogen molecule. As with the QAOA testing, noise profiles from HF (ibmq\_kolkata) and LF (ibmq\_toronto) devices were used. As shown in Figure~\ref{fig:vqe_results}, \framework{} achieves a ground state energy within 0.3\% of the HF-only optimization. By dynamically allocating executions across the HF and LF devices, \framework{} introduces almost no additional executions beyond those required for HF or LF alone. \framework{} matches the accuracy of an HF optimization while reducing the computational load on that device. This highlights the benefits of using \framework{}.

\subsection{Case Study: Asynchronous Gradient Descent}
\label{sec:eqc}
\begin{figure}[h!]
    \centering
    \includegraphics[width=0.82\linewidth]{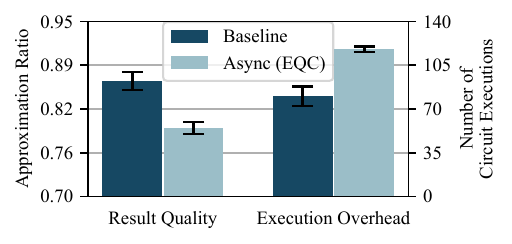}
    \caption{Accuracy and executions after one epoch of asynchronous gradient descent (AGD) on a 3-layer QAOA circuit. AGD requires more executions than optimizing all parameters together and achieves lower accuracy.}
    \label{fig:agd_compare}
\end{figure}

We compare our \framework{} against the asynchronous gradient descent (AGD) optimization used in EQC~\cite{stein2022eqc}. EQC optimizes individual parameters on different devices separately, and the results are combined at each epoch's end. \framework{} differs by optimizing all parameters together across multiple devices. We evaluate both approaches using a 3-layer QAOA circuit. Figure~\ref{fig:agd_compare} shows that even a single AGD epoch requires more executions than optimizing all parameters together on the HF device. Also, the approximation ratio after 1 AGD epoch is much lower than \framework{} or HF only. EQC results also show slower convergence when distributing optimizations across multiple devices compared to using only a high-fidelity device due to the limitations of the low-fidelity devices. In contrast, \framework{} is not constrained by averaging intermediate results and fully leverages both devices throughout the execution. Overall, the synchronous optimization approach in \framework{} is better suited for leveraging multiple devices than AGD. 

\section{Related Works}

\subsection{Improving Utilization In Quantum Clouds}

Improving the utilization of devices in quantum clouds has been previously studied. Recent works have emphasized multiprogramming to enhance throughput and resource utilization. For instance, Das et al.~\cite{das2019case} propose concurrent execution of multiple quantum workloads on NISQ machines. Building on this concept, Resch et al.~\cite{resch2021accelerating} extend the idea to study circuit-level concurrency specifically for VQA circuits. On the other hand, QuCloud~\cite{liu2021qucloud} focuses on optimizing qubit mapping to efficiently allocate quantum resources among various programs in a cloud environment.
Similarly, CutQC~\cite{tang2021cutqc} introduces a hybrid classical-quantum computing method, partitioning large quantum circuits to execute on smaller quantum devices. These approaches, focusing on throughput enhancement, are orthogonal to \framework{}, which explores leveraging different quantum devices at various optimization stages of VQAs. Combining these methodologies could lead to more effective and efficient quantum cloud computing paradigms.

\subsection{VQA Optimization}
Training VQAs poses substantial execution overheads. Prior work highlights a clustering phenomenon in QAOA~\cite{brandao2018fixed}, indicating that optimal parameters from one problem can be transferred to others, providing near-optimal outcomes~\cite{galda2021transferability, shaydulin2023parameter}. Conversely, CAFQA~\cite{ravi2022cafqa} employs classical simulation to optimize VQE, using a Clifford circuit variant that is classically simulatable. However, this type of warm start initialization does not apply to QAOA~\cite{cain2023qaoa}. Ansatz optimization is also crucial for enhancing algorithmic efficiency. AdaptVQE~\cite{grimsley2019adaptive}, for instance, dynamically constructs the ansatz during the execution to minimize the depth of the ansatz circuit.

\subsection{Hybrid Optimization}
In classical optimization, an extensive set of studies focused on hybrid models that strategically combine different techniques to improve performance. Prior work~\cite{gadzinski2022combining} proposes a technique merging fast and frugal decision trees with machine learning models. This hybrid approach aims to couple the interpretability of trees with the superior accuracy of neural networks. Similarly, Kudva et al.~\cite{kudva2022constrained} investigate constrained Bayesian optimization algorithms for tuning noisy black-box functions. By incorporating robust stochastic models within an efficient sequential design strategy, their method provides regret bounds unmatched by either paradigm alone. Beyond marrying simpler and more sophisticated techniques, researchers also envision hybrids of different cutting-edge approaches. He et al.~\cite{he2023review} review recent work on surrogate-assisted evolutionary algorithms capable of optimizing extremely expensive objective functions. Gaussian process-based Bayesian optimization has proven highly effective at handling single and multi-objective design problems involving complex systems.  If these classical optimization methods are suitable for the exploration phase of VQA optimization, they could be integrated with \framework{}, as most optimization approaches comprise initial exploration followed by fine-tuning.

\section{Conclusion}
Quantum cloud services are essential for enabling remote access to quantum devices, but they struggle with a significant demand-supply mismatch and uneven load distribution due to varying device fidelity. As a result, program execution is either impacted by the wait times of the high-fidelity devices or the noise of the low-fidelity devices. This paper proposes {\em \framework{}}, a fully automated job scheduling framework that splits the program execution into phases with varying noise resilience. \framework{} schedules the more noise-resilient phase on the low-fidelity device (for higher execution throughput). It runs the more noise-sensitive phase on the high-fidelity device (for higher application fidelity). \framework{} also leverages the insight that near-term quantum applications comprise several independent training rounds that start from different initial points, but not all of them lead to the optimal solution. \framework{} quickly evaluates these candidates on the low-latency device, eliminates the weaker candidates, and runs the remaining ones on the high-fidelity machine. Overall, \framework{} offers similar solutions 17.4 $\times$ faster than the baseline, or it provides 13.3\% better solutions on average with a similar execution time.

\section*{Ackowledgement}
We would like to thank the anonymous reviewers. We would also like to thank Yunong Shi for his insightful comments, discussions, and feedback on our work. This work was supported by the National Research Council (NRC) Canada grants AQC 003 and AQC 213, as well as the Natural Sciences and Engineering Research Council of Canada (NSERC) [funding number RGPIN-2019-05059]. This research used the National Energy Research Scientific Computing Center (NERSC) resources, a U.S. Department of Energy Office of Science User Facility located at Lawrence Berkeley National Laboratory, operated under Contract No. DE-AC02-05CH11231. Poulami Das acknowledges the generous support through the AMD endowment at UT Austin. The views contained herein are those of the authors and should not be interpreted as endorsements of NSERC, NRC, UT Austin, or UBC.

\balance
\begin{appendices}
\renewcommand{\thesectiondis}[2]{\Alph{section}:}

\section{Artifact Appendix}

\subsection{Abstract}
Our artifacts include the Qoncord job scheduling framework for Variational Quantum Algorithms (VQAs), designed for cloud-based Noisy Intermediate-Scale Quantum (NISQ) systems but primarily tested using classical noisy quantum circuit simulators. Qoncord optimizes the execution of VQAs by strategically dividing the training process into exploratory and fine-tuning phases, and distributing these across different quantum devices based on their fidelity and availability. The framework includes algorithms for adaptive scheduling and optimization of restart procedures. Our artifacts demonstrate Qoncord's ability to achieve solutions 17.4× faster than baseline approaches and deliver 13.3\% better solutions within the same time budget using simulated NISQ environments.

\subsection{Artifact check-list (meta-information)}
{\small
\begin{itemize}
\item {\bf Hardware: } Classical simulation can be running on both CPU and GPU platforms
\item {\bf Metrics: } Execution overhead, load balancing, solution quality
\item {\bf Output: } Optimized VQA (QAOA and VQE) solutions from simulated quantum environments
\item {\bf Experiments: } Comparison of Qoncord vs. baseline scheduling approaches using quantum circuit simulators for QAOA and VQE problems
\item {\bf How much time is needed to prepare workflow (approximately)?: } Less than 10min
\item {\bf How much time is needed to complete experiments (approximately)?: } About 30 minutes
\item {\bf Publicly available?: } Yes
\item {\bf Code licenses (if publicly available)?: } MIT license
\item {\bf Workflow automation framework used?: } No
\item {\bf Archived (provide DOI)?: } N/A
\end{itemize}
}

\subsection{Description}

\subsubsection{How to access}

The Qoncord framework and associated code for our experiments are publicly available via GitHub at: https://github.com/meng-ubc/Qoncord

\subsubsection{Hardware dependencies}
Our experiments were primarily conducted using classical hardware to simulate noisy quantum circuits. While Qoncord is designed for cloud-based NISQ systems, no actual quantum hardware is required to run our simulations. There are no strict hardware requirements for running the code, but for optimal execution time, we recommend at least the following setup:

\begin{itemize}
    \item \textbf{CPU}: 8-core processor
    \item \textbf{RAM}: 16 GB
    \item \textbf{GPU}: Recommended for noisy simulation experiments with more than 10 qubits
    \item \textbf{Storage}: No particular requirements; the code and data files occupy less than tens of megabytes
\end{itemize}

Note that while these specifications are recommended for better performance, the code can run on less powerful machines, which will take longer execution times.

\subsubsection{Software dependencies}
To run our code and reproduce the experiment results presented in the paper, the following software dependencies are required:

\begin{itemize}
\item \textbf{Quantum Computing}: qiskit, qiskit-ibm-runtime, qiskit-aer
\item \textbf{Data Visualization}: matplotlib, seaborn
\item \textbf{Additional Libraries}: numpy, scipy, networkx, tqdm
\end{itemize}

A complete list of dependencies with specific versions that have been tested can be found in the \texttt{requirements.txt} file in our repository.

\subsubsection{Data sets}
Our experiments use two types of VQAs: QAOA with randomly generated NetworkX graphs, and VQE with the quantum observable of a hydrogen molecule using the UCCSD ansatz. Detailed information about these can be found in the methodology section.

For the noise model, we utilized the fake backend feature from Qiskit, which incorporates noise profiles that were profiled on real quantum devices.

\subsection{Installation}

To install Qoncord and its dependencies, follow these steps:

\begin{enumerate}
    \item Clone the repository:
    \begin{verbatim}
    git clone https://github.com/ \
    meng-ubc/Qoncord.git
    cd Qoncord
    \end{verbatim}

    \item Create a virtual environment (optional but recommended):
    \begin{verbatim}
    conda create -n qoncord_env
    conda activate qoncord_env
    \end{verbatim}

    \item Install the required dependencies:
    \begin{verbatim}
    pip install -r requirements.txt
    \end{verbatim}
\end{enumerate}

\subsection{Experiment workflow}
To facilitate the reproduction of our key results, we have prepared a Python script for each experiment. Running any of these scripts is straightforward—simply execute the command \texttt{python script.py}. This will automatically run the corresponding experiment and save the results to a file. Then a plotting script can be used to generate the relevant plots.

Each figure produced by these scripts can be directly compared against the figures presented in the paper, ensuring an easy and accurate validation of the results.

\subsection{Evaluation and Expected Results}

This section provides a detailed evaluation of each experiment. Each subsubsection introduces the experiment, specifies the Python script to be executed, identifies the corresponding results in the paper, outlines the expected runtime, and briefly explains how to interpret the output.

\subsubsection{\textbf{Queue Simulation}}

This experiment simulates a quantum queue to evaluate different scheduling policies under various VQA job ratios. The goal is to analyze the trade-offs between execution fidelity and throughput.

\textbf{Script:}
\texttt{main\_queue\_sim.py}

\textbf{Paper Results:} Section VI.A, specifically Figure 12.

\textbf{Expected Runtime:}
Approximately 2 minutes.

\textbf{Expected Results:}
A scatter plot showing execution fidelity versus throughput for all scheduling policies and VQA job ratios. This plot can be directly compared to Figure 12.

\subsubsection{\textbf{Multi-Restart QAOA Optimization}}

This experiment evaluates the performance of different multi-restart strategies for QAOA optimization and compares their circuit execution overhead across various quantum devices.

\textbf{Script:}
\texttt{2\_qaoa\_optimization.py}

\textbf{Paper Results:} Section VI.B, specifically Figures 13, 14.

\textbf{Expected Runtime:}
Approximately 10 minutes.

\textbf{Expected Results:} A box plot showing the distribution of the approximation ratio for different approaches. Also, a bar plot displaying the circuit execution overhead on each device.
These plots correspond to Figures 13 and 14 in the paper.

\subsubsection{\textbf{Single-Restart VQE Optimization}}

This experiment focuses on optimizing VQE using a single-restart approach, comparing the performance and execution overhead of different execution modes.

\textbf{Script:}
\texttt{3\_vqe\_optimization.py}

\textbf{Paper Results:} Section VI.F, specifically Figure 21.

\textbf{Expected Runtime:}
Approximately 4 minutes.

\textbf{Expected Results:}
A bar plot showing the approximation ratio and circuit execution overhead for different execution modes, which can be compared to Figure 21 in the paper.

\subsubsection{\textbf{Comparison to Asynchronous Gradient Descent}}

This experiment compares the performance of the default gradient descent method with an asynchronous gradient descent approach in optimizing quantum circuits.

\textbf{Script:}
\texttt{4\_ae\_async.py}

\textbf{Paper Results:} Section VI.G, specifically Figure 22.

\textbf{Expected Runtime:}
Approximately 6 minutes.

\textbf{Expected Results:}
A bar plot showing the approximation ratio and circuit execution overhead for both default and asynchronous gradient descent methods.

\end{appendices}

\bibliographystyle{IEEEtran}
\bibliography{refs}

\end{document}